\documentclass[12pt,a4paper]{article}
\usepackage[english]{babel}
\usepackage{hyperref}
\usepackage{amsthm}
\usepackage{amssymb,amsmath}
\usepackage{euler}
\usepackage{listings}
\usepackage[cm]{fullpage}
\usepackage{makeidx}
\makeindex

\title{An implementation of range trees with fractional cascading in C++}

\author{Vissarion Fisikopoulos}

\theoremstyle{remark}

\theoremstyle{definition}

\begin{document}

\maketitle

\abstract{Range trees are multidimensional binary trees
which are used to perform d-dimensional orthogonal range searching. In this
technical report we study the implementation issues of range trees with
fractional cascading, named layered range trees. We also document our
implementation of range trees with fractional cascading in C++ using STL and
generic programming techniques.}

\section{Introduction}
This project is an implementation of range trees with fractional cascading, named layered range trees in C++ using STL and generic programming techniques. Range trees are multidimensional binary trees which are used to perform d-dimensional orthogonal range searching. Range trees were discovered independently by several people including Bentley\cite{361007}, who also discovered kd-trees and Lueker, who introduces the technique of fractional cascading for range trees \cite{1382567}. An introduction to orthogonal range searching, range trees and fractional cascading can be found in \cite{compgeom:2000, lecnotes}. In \cite{implement} there is a presentation of a project of efficient implementations of range trees in 2-3 dimensions including layered ones and some experimental results.

\section{Complexity issues}
The range trees answer a d-dimensional range query in time $O(\log^{d} n + k)$, where $n$ is the whole set of points and $k$ is the set of reported points. The construction time and the space the tree consume are $O(n\log^{d-1} n)$. Using fractional cascading we can be benefited by a $\log n$ factor in the last level of the tree and the resulting time complexity become $O(\log^{d-1} n + k)$. Intuitively, fractional cascading perform one binary search instead of two in the last level. The optimal solution to the orthogonal range search problem is due to Chazelle \cite{79149,77614} who propose a structure with time complexity $O(\log^{c} n + k)$ and $O(n(\log n/\log\log n)^{d-1})$ space consumption, where $c$ is a constant. 

\section{Range trees in CGAL}
Although, CGAL library \cite{cgal:eb-08} provides some classes for range trees there is space for optimizations in that package \cite{neyer}. Firstly, there is a lack of recursive construction of d-dimensional range tree and the only way to construct a range tree of dimension d is to build a tree of dimension 1 and then make this an associative range tree of a new one which will have dimension 2. Then one must build a tree of dimension 3 with this tree as an associative tree and this technique continues until the construction of the whole d-dimensional tree.

In addition to that, the package uses virtual functions, which increases the run time and finally there is no fractional cascading.

The proposed approach uses nested templates for the representation of the $d$-dimensional range tree which is defined in compilation time. The dimension of the tree must be a constant and defined in the compilation time. In the last level a fractional cascading structure is constructed. 

For example a 4-dimensional range tree of size $n$ with different kind of data at each layer is given by the following nested templated definition.
\vspace{.5cm}
\lstset{language=c++}
\lstset{commentstyle=\textit}
\begin{lstlisting}[frame=trbl]{}
Layered_range_tree <DataClass,
                  Layered_range_tree <DataClass,
                                    Last_range_tree <DataClass>
                                     >
                  >  tree(n);
\end{lstlisting}
\vspace{.5cm}

Note that for each layer $i<d-1$ the same class \texttt{Layered\_range\_tree} is used. The last two layers, in which the fractional cascading is implemented,
use the \texttt{Last\_range\_tree} class. The DataClass has the definitions of
each layer's own data along with the comparison operators. 

\section{Software implementation}
Essentially, the project was implemented using the C++ language and the STL library\cite{citeulike:1120270}. Concisely, the design uses methods from object oriented as well as the generic programming style. 

\paragraph{Representation.} The trees are represented as STL vectors. The tree
traversals are implemented using index arithmetic i.e. node's $i$ parent is
$\lfloor i/2\rfloor$, the left, right child of $i$ is $2i+1$ and $2i+2$
respectively. This method is optimal for a \textit{full, static, binary} tree
and in our case the third is always hold. In order to have a full binary tree we
replicate the last (biggest in the fist dimension) point and in the worst case
we have a tree the half of which is useless with no effect to the time
complexity (the replicated nodes would not be visited). In this project we are
interested in the static case of range trees but the design is sufficient for a
dynamic implementation in which the tree nodes must also have some extra
pointers. On the other hand, dynamization of the fractional cascading structure
is not trivial \cite{ALGOR::MehlhornN1990}.

\paragraph{Construction.} For the construction of the tree we need to sort the input data with respect to the first coordinate and build recursively (top-down) the main tree in linear time. For the associative trees we don't have to sort the input data again. We build the associative trees in bottom-up manner. Every node merge the sorted lists of its children in linear time starting from the leaves which are trivially sorted. Note that this is essentially the same algorithm as merge-sort. 

\paragraph{Memory consumption.} Even  the asymptotic complexity of space stated
above ensures that range tree needs a lot of memory. The only constraint in the
number of dimensions of data is memory. Moreover, from the asymptotic complexity
follows that with fixed memory there is a trade of between the number of data
and number of dimensions. For example see
\begin{center}
\href{
http://users.uoa.gr/~vfisikop/compgeom/Layered_Range_trees/examples/Layered_R
ange_tree_10.cpp}{
\scriptsize
http://users.uoa.gr/$\sim$vfisikop/compgeom/Layered\_Range\_trees/examples/Layer
ed\_Range\_tree\_10.cpp}
\end{center}
for an example of a range tree over 10-dimensional data. 

\section{Future work}  The most important feature for improvement is in the
case that the tree is not full. The solution proposed in the previous section, 
seams very suboptimal and is not implemented yet. Another point is that the
implementation don't handle yet the case that two points have equal coordinates.

\section{Acknowledgements} This project started as a course project in the
graduate course of Computational Geometry, National and Kapodistrian University
of Athens 2008-2009 under the supervision of professor Ioannis Z. Emiris. The
C++ code
can be
found in:
\begin{center}
\href{http://cgi.di.uoa.gr/~vfisikop/compgeom/Layered_Range_trees/}{
http://cgi.di.uoa.gr/$\sim$vfisikop/compgeom/Layered\_Range\_trees}             
\end{center}

\bibliographystyle{abbrv}
\bibliography{references}

\end{document}